\documentclass[10pt,twocolumn,letterpaper]{article}

\usepackage{wacv}
\usepackage{times}
\usepackage{epsfig}
\usepackage{graphicx}
\usepackage{amsmath}
\usepackage{amssymb}
\usepackage{hyperref}

\wacvfinalcopy 
\def\wacvPaperID{54} 

\ifwacvfinal\fi
\begin{document}

\title{Multi-FAN: Multi-Spectral Mosaic Super-Resolution Via Multi-Scale Feature Aggregation Network}

\author{Mehrdad Shoeiby \\
CSIRO-DATA61\\
{\tt\small mehrdad.shoeiby@data61.csiro.au}
\and
Sadegh Aliakbarian \\
ANU/CSIRO-DATA61\\
\and
Saeed Anwar \\
CSIRO-DATA61\\
\and
Lars Petersson \\
CSIRO-DATA61\\
}

\maketitle
\ifwacvfinal\thispagestyle{empty}\fi

\begin{abstract}
This paper introduces a novel method to super-resolve multi-spectral images captured by modern real-time single-shot mosaic image sensors, also known as multi-spectral cameras. Our contribution is two-fold. Firstly, we super-resolve multi-spectral images from mosaic images rather than image cubes, which helps to take into account the spatial offset of each wavelength. Secondly, we introduce an external multi-scale feature aggregation network (Multi-FAN) which concatenates the feature maps with different levels of semantic information throughout a super-resolution (SR) network. A cascade of convolutional layers then implicitly selects the most valuable feature maps to generate a mosaic image. This mosaic image is then merged with the mosaic image generated by the SR network to produce a quantitatively superior image. We apply our Multi-FAN to RCAN (Residual Channel Attention Network), which is the state-of-the-art SR algorithm. We show that Multi-FAN improves both quantitative results and well as inference time.
\end{abstract}

\section{Introduction}
\label{sec:intro}
 Recent development of real-time snapshot mosaic image sensors have given rise to fast and portable multi-spectral camera devices,  with performance comparable to modern trichromatic (RGB) cameras~\cite{cdd_hyper_2013,cmos_review_2006}. Despite the great interest in these multi-spectral imaging devices, with applications ranging from astronomy~\cite{msi_astronomy_2016} to object detection in autonomous vehicles \cite{takumi_2017,lars_2013}, they suffer from an inherent constraint, that is, a trade-off between the spatial and the spectral resolution. The reason is that there is a limited physical space on a 2D camera image sensor. A higher spatial resolution (smaller pixel size) reduces the number of wavelength bins that can fit on that pixel on the image sensor. This creates a limitation in certain applications where size and portability are essential factors, for instance, on a UAV~\cite{doering_2016}. A more portable (smaller/lighter) camera results in a device that suffers more from lower spatial and spectral resolution. 

Despite the fact that multi-spectral cameras suffer from low spatial resolution compared to their RGB counterparts, and considering the vast interest in these devices, there is a lack of scientific literature on multi-spectral image super-resolution (SR). Therefore, in this article, we aim to develop a SR model for multi-spectral cameras, \ie, SR of the mosaic image sensors, to address the gap identified in the literature. 
However, it should be noted that our method is general and robust as it can be applied to any mosaic snapshot sensor images.  In this paper, we claim the following contributions:

\begin{figure*}
    \centering
    \includegraphics[width=\textwidth]{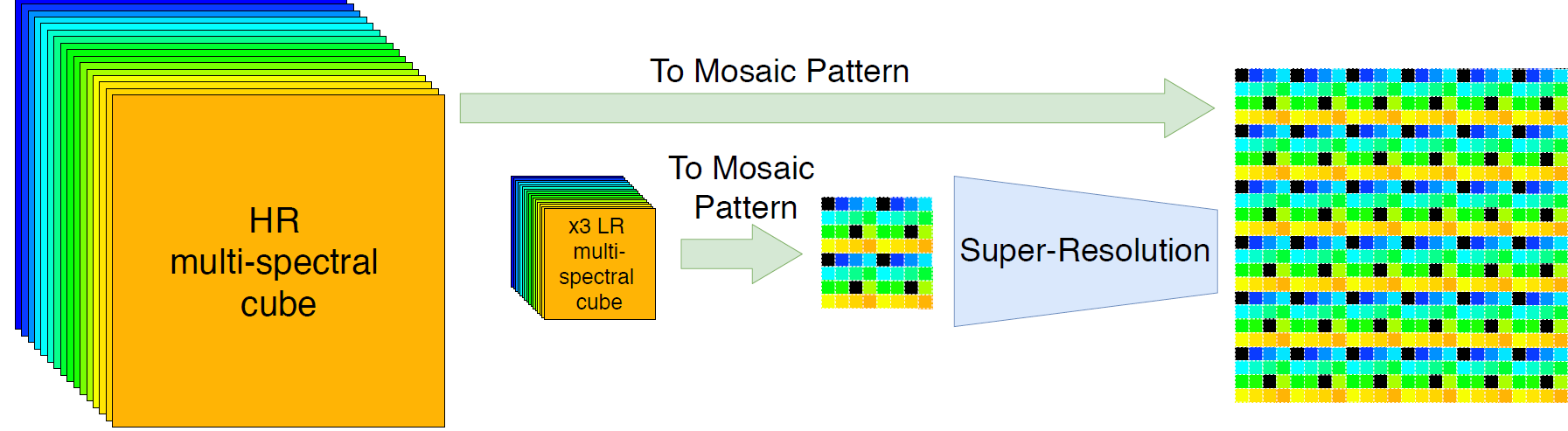}
    \caption{Illustration of mosaic pattern generation procedure. Given the HR and LR cubes, we generate mosaic pattern images according to their wavelength position on the image sensor  depicted in \cite{shoeiby_dataset_2018}, which are then used to train our super-resolution network.}
    \label{fig:mosaic}
\end{figure*}

\paragraph*{Contributions}
\begin{itemize}
    \item We propose a multi-spectral SR method that enhances the state of the art RCAN network in terms of performance and inference time.
    \item A simple, yet novel, data representation, which exploits the mosaic pattern of a snapshot sensor, that allows super-resolving from the mosaic image rather image cubes which is customary in the SR literature. Our work is, to the best of our knowledge, the first CNN-based technique to super-resolve multi-spectral images from mosaic images directly.
\end{itemize}

 \section{Related Work}
 \label{section:relate_work}
Here, we provide details about the current state-of-the-art RGB and multi-spectral SR algorithms. Dong~\etal~\cite{dong2016SRCNNPAMI} furnishes the pioneering work in CNN SR. The proposed linear network coined super-resolution convolution neural network (SRCNN)~\cite{dong2016SRCNNPAMI} is composed of three convolutional layers only, with various filter sizes. SRCNN~\cite{dong2016SRCNNPAMI} improved significantly over the traditional SR methods. Due to the success of SRCNN, many CNN based algorithms~\cite{lim2017EDSR,dong2016FSRCNN,zhang2018SRMDNF} were proposed. 

Initially, the focus was on simple linear networks~\cite{dong2016FSRCNN,zhang2018SRMDNF}, \ie, networks without any skip-connections and learned the image itself. Although comprised of eight convolutional layers, fast super-resolution convolutional neural network, abbreviated as FSRCNN~\cite{dong2016FSRCNN} speeds up the SR process by using as input the original low-resolution patch instead of a bi-cubically up-sampled one highlighting the fact that using interpolation to scale up images deteriorates SR performance. Recently, another linear network, SRMD which stands for super-resolution network for multiple degradations ~\cite{zhang2018SRMDNF}, is proposed, which can handle multiple degradations \ie deblurring, and unknown downscaling operators. 

Other approaches use residual connections~\cite{kim2016VDSR,lim2017EDSR,ahn2018CARN}. For example, Kim~\etal~\cite{kim2016VDSR} introduced very deep super-resolution (VDSR), which has a single global skip-connection from the input to the final output. Similarly, enhanced deep super-resolution \ie EDSR~\cite{lim2017EDSR} employs residual blocks (RBs). More recently, cascading residual network (CARN)~\cite{ahn2018CARN} is proposed, which also employs a variant of RBs with cascading connections. CARN~\cite{ahn2018CARN} lags behind EDSR~\cite{lim2017EDSR} in terms of PSNR; however, its focus is on efficiency and speed.

With the rise of the usage of skip-connections, Kim~\etal~\cite{kim2016DRCN} utilized recursive connections, namely, deep recursive convolutional network (DRCN), which applies the same convolutions' several times, as this helps in keeping the number of parameters small. Similar to~\cite{kim2016DRCN}, deep recursive residual network (DRRN)~\cite{tai2017DRRN} proposed a deeper network replicating the necessary skip-connection modules many times to realize a multi-path architecture. Furthermore, a persistent memory network for image superresolution (abbreviated as MemNet)~\cite{tai2017memnet} relies on a recursive block that is like the RBs defined in~\cite{lim2017EDSR}. The performances of the recursive skip-connection networks are comparable to each other. 

More lately, motivated by the success of DenseNet~\cite{huang2017densely}, CNN-based SR networks concentrated on the dense connection model. For example, SRDenseNet~\cite{tong2017image} uses dense connections to learn compact models, avoiding the problem of vanishing gradients and ease the flow from low-level features to high-level features. The authors of~\cite{tong2017image} proposed a sequential arrangement of the dense modules followed by deconvolutional layers to reconstruct the final output from the high-level features only. Recently, residual-dense network (RDN)~\cite{zhang2018RDN} employed dense connections to learn the local representations from the patches at hand. Likewise, the dense-deep back-projection network, also known as DDBPN~\cite{haris2018DDBPN} aims to model a feedback mechanism with a feed-forward procedure. Hence, they proposed a series of densely connected upsampling and downsampling layers and combined the intermediate outputs to predict the resulting SR image.   

To improve the perceptual quality of the images, SR networks based on GANs~\cite{radford2015supervisedGAN,goodfellow2014generative} have been attempted. Interesting work in this regard is SRResNet~\cite{ledig2017SRGAN} that combine three different losses, \ie, $\ell_2$, perceptual, and adversarial. The generator is composed of RBs and long skip-connections. The SRResNet outperformed its competitors in terms of perceived quality by a significant margin. More recently, SRFeat~\cite{park2018srfeat} was proposed, which uses an additional discriminator to assist the generator in creating more realistic images. ESRGAN~\cite{wang2018esrgan} is inspired by~\cite{ledig2017SRGAN} where batch normalization is removed, and dense connections between five consecutive convolutional layers are incorporated. Furthermore, to enforce residual learning, ESRGAN~\cite{wang2018esrgan} also has a global skip-connection. Moreover, instead of using the traditional discriminator, ESRGAN~\cite{wang2018esrgan} exercises an enhanced discriminator called Relativistic GAN~\cite{jolicoeur2018RGAN}. 

The visual attention~\cite{mnih2014recurrent} concept is brought to SR by RCAN~\cite{zhang2018RCAN}, which models the inter-channel dependencies using a channel attention (CA)mechanism. This process is coupled with a very deep network composed of groups of RBs. Following in the footsteps of~\cite{zhang2018RCAN}, the residual attention module (SRRAM) by~\cite{kim2018ram}, employed a dual attention approach, \ie, both spatial attention and CA to model the inter-channel and the intra-channel dependencies, respectively; however, SRRAM~\cite{kim2018ram} lags behind RCAN~\cite{zhang2018RCAN}.  

(SRRAM) by~\cite{kim2018ram}, employed a dual attention approach, \ie, both spatial attention and channel attention to model the inter-channel and the intra-channel dependencies, respectively; however, SRRAM~\cite{kim2018ram} lags behind RCAN~\cite{zhang2018RCAN} and DRLN~\cite{anwar2019densely}. 


The SR algorithms mentioned above focus mainly on super-resolving RGB images even though the multi-spectral images are comparatively more adversely affected by the resolution constraints. These constraints are due to the inherent characteristics of the multi-spectral cameras. Moreover, current SR algorithms are designed for demosaiced images, and not for mosaic/cube images requiring them to take into account spectral and spatial correlation of multiple channels. The scarcity of multi-spectral SR algorithms may be due to the absence of multi-spectral SR benchmarking platforms as well as the difficulty of accessing suitable SR spectral datasets. For example,  Li~\etal~\cite{li_sr_spectral_2017} aims to improve the quality of hyperspectral (not multi-spectral) images and is one of the few CNN spectral SR methods.  To the best of our knowledge, the only multi-spectral SR methods~\cite{lahoud_2018,shi_2018}, were submitted to the PIRM2018 multi-spectral SR challenge~\cite{shoeiby_pirm2018_method,shoeiby_dataset_2018}. To super-resolve the images, Lahoud~\etal~\cite{lahoud_2018} adopted an image completion technique that requires $\times 2$ and $\times 3$ down-sampled images as input to a 12 layer convolutional network. While achieving good results, it addresses the problem of $\times 3$ SR given $\times 2$ and $\times 3$ down-sampled images, rather than single image SR. It is also not an end-to-end CNN based implementation. The best end-to-end CNN based method in the challenge was proposed by Shi~\etal~\cite{shi_2018}, which implicitly employed the RCAN~\cite{zhang2018RCAN}.  

All the methods in the challenge do not directly take into account any spectral correlation or consider the spatial offsets of each wavelength channel. These methods only rely on the network to learn the inherent geometrical relationship of each pixel; to elaborate further, the input data encompass multi-spectral image cubes, which is a customary way to represent multi-spectral images. However, unlike hyperspectral cubes, where each pixel location contains information for the complete measurable spectral range, with multi-spectral mosaic sensors, each raw pixel location only contains information about one specific wavelength band. In other words, representing a mosaic image in the form of a multi-spectral cube implies that the precise geometrical information inherent to each raw pixel is lost during the process of conversion. 

Preserving the aforementioned geometrical information can be achieved by creating the multi-spectral cubes via interpolation or demosaicing method \cite{monno2015practical,jaiswal2016adaptive,dong2016FSRCNN,zhang_2018} to the dimension of the original mosaic image. However, demosaicing or interpolation has shown to deteriorate SR performance \cite{fu_2018,zhou_2018,shoeiby_color-predict_2019}. A couple of related works exploited mosaic images, instead of interpolation/demosaicing include Fu~\etal~\cite{fu_2018} which utilizes a variational approach that exploits the mosaic RGB images to super-resolve hyperspectral images. This is preceded by Zhou~\etal~\cite{zhou_2018}, which presented a deep residual network for super-resolving RGB that utilizes the generated (estimated) mosaic images. Both works highlight the fact that demosaicing involves interpolation (\eg, bicubic) and introduce artifacts that can deteriorate the performance of the algorithms.  

\section{Proposed Method}
\begin{figure*}
  \includegraphics[width=\linewidth]{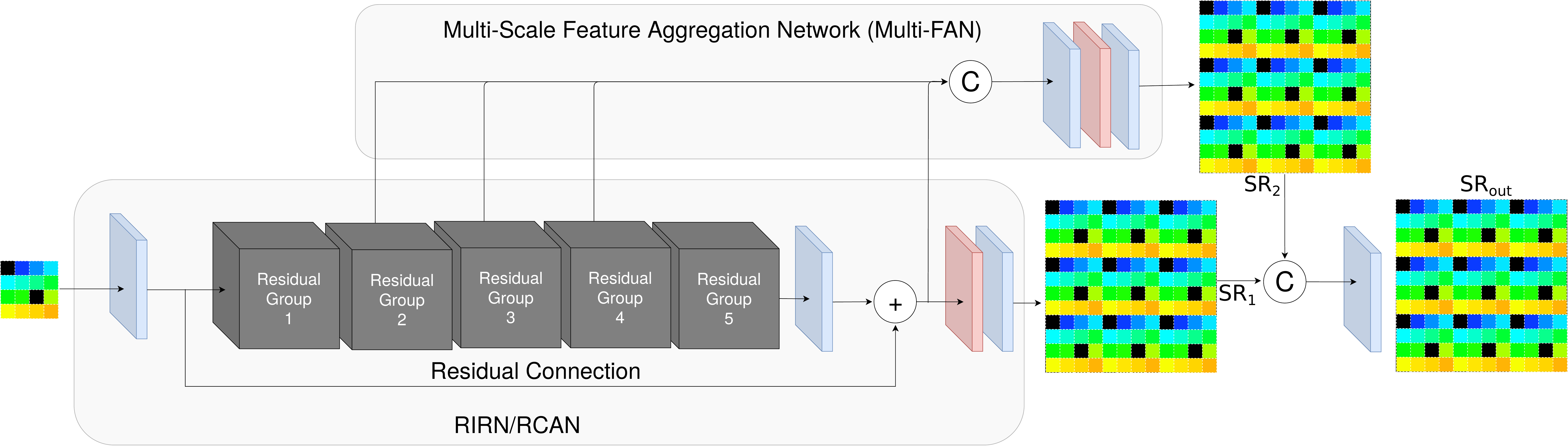}
  \caption{Overview of our propose SR network. The model consists of two main blocks: RIRN/RCAN and Multi-FAN. While RIRN/RCAN attempts to super-resolve the input given the addition of the output of the first convolutional layer and the last residual group (i.e.,  RG$_5$), Multi-FAN aims to fuse features at multiple levels (including the ones from shallower RGs) to perform SR. The final output is then computed given the information derived from these two networks. Note, there are $g=5$ RGs $b=3$ RBs for RCAN which corresponds to the network of~\cite{shi_2018}. Channel-wise concatenation is denoted by {\large{\copyright}}. 2D convolutional layers, and 2D transposed convolutional layers are denoted by blue and red blocks respectively.}
  \label{fig:network}
\end{figure*}

As explained in section \ref{section:relate_work}, multi-spectral cubes provided in the PIRM2018 challenge do not contain the geometric information of each specific wavelength. To introduce this geometric information into the dataset without having to interpolate each channel, we propose the generation of mosaic LR and HR from raw multi-spectral cubes to carry out SR on mosaic images rather than image cubes. Moreover, we propose to utilize a Multi-FAN mechanism on top of features of RCAN~\cite{shi_2018,zhang_2018} at multiple semantic scales. This allows the model to effectively consider high-level semantic features (with larger receptive fields), as well as the ones from shallower layers, which capture more details (e.g., textures and edges, and multi-spectral intra-channel dependencies) but in local regions of the input images (with smaller receptive fields). The final super-resolved image is then computed given the information from multiple semantic scales and information from the deepest convolution layer. In addition, we empirically observed a trade-off between the effectiveness and the inference time of the CA mechanism. Hence, we also propose that removing CAs and using our Multi-FAN can improve inference time significantly while improving/maintaining PSNR performance.

\subsection{Residual in Residual Channel Attention Network (RCAN)}
In our experiments, we use RCAN, which is state-of-the-art in SR as our baseline. Since we also study the effect of the CA mechanism, we refer to the baseline as RCAN or RIRN (RCAN without CA) where appropriate. The RCAN/RIRN network block in Figure \ref{fig:network} is similar in structure to the RCAN in \cite{shi_2018, zhang_2018}, and in particular, the RCAN structure used is \cite{shi_2018} for multi-spectral SR using the same dataset. The body is comprised of $g$ number of sequential residual groups (RGs). Each RG contains $b$ number of residual blocks (RBs). We choose $g=5$, $b=3$ to be consistent with the parameters of RCAN in \cite{shi_2018}. The tail of the network constitutes a 2D convolution layer creating $C$ feature channels from the input LR mosaic image. The head of the RIRN network is the reconstruction part, which consists of a transpose convolutional layer to up-sample the features and followed by a convolutional layer to produce the HR mosaic image. The network, given the input multi-spectral ($I_{LR_{MS}}$) and the high-resolution ground truth image ($I_{HR_{MS}}$), is trained to super-resolve the input image.


\subsection{Multi-scale Feature Aggregation Network (Multi-FAN)}

Through a general observation of multi-spectral pixels, one could intuitively infer/observe a general relationship between the wavelengths corresponding to a single multi-spectral pixel \cite{spectral_similarity_2000}. As can be seen in Figures \ref{fig:mosaic} and \ref{fig:network}, each multi-spectral pixel is a $4\time4$ filter pattern capturing 16 different wavelengths. These wavelengths contain information about the material in the scene and their relationship, and considering the $4\time4$ pattern of a multi-spectral pixel, shallower semantic features could contain this intra-channel dependency information.

As illustrated in Figure~\ref{fig:network}, the RCAN/RIRN output ($SR_{1}$) is generated after the residual connection; thus, the output is produced given the addition of the output of the first convolutional layer and the last RG (\ie, RG$_5$). While such features are rich enough to super-resolve the input image, we argue that lower level features, \ie, the ones from shallower RGs, account for fine-grained details, such as edges and texture, and possibly, as explained earlier, the intra-wavelength dependencies which are essential to consider when performing SR. To achieve this within our network, we concatenate the output of the last three RGs (RG$_2$ to RG$_{g-1}$) in the network with the final feature map, right before the transposed convolutional layer in RIRN. The three feature maps from RG$_2$ to RG$_{g-2}$ (RG$_4$) contain valuable information of the various level of semantic features with different receptive fields. The final feature map, thanks to the skip-connection, contains very low-level semantic information via smaller receptive fields as well as the highest level semantic information with wider receptive fields.  

After concatenating the four feature maps, we have a feature map with a larger number of channels ($4\times C$ ). To extract the most valuable information from this feature representation and to facilitate feature reduction, we feed the $4\times C$ feature channels through a cascade of a 2D convolutional layer, a ReLU activation layer, a 2D transposed convolutional layer, a ReLU activation layer, followed by a 2D convolution layer we produce a mosaic image ($SR_{2}$). This mosaic image is then concatenated with the mosaic image produced by RCAN/RIRN ($SR_{1}$) and a 2D convolution layer learns to merge the two super-resolved images that leads to a lower value for the cost function and a quantitatively superior image ($SR_{out}$).

\subsection{Loss functions}

In the CNN-based SR literature, a simple loss function such as $L_1$ \cite{zhang_2018} or $L_2$ \cite{shi_2018,lahoud_2018}, or a perceptual loss function such as SSIM \cite{ssim_wang_2004} is usually utilized to train models. Here, for consistency, we choose $L_1$ loss as our baseline loss function since an $L_1$ function is less sensitive to outliers compared to an $L_2$ function. For our baseline $L_1$ function, we use the $SmoothL_1$ \cite{smoothl1} PyTorch\cite{pytorch} implementation, which is a more stable implementation compared to vanilla $L_1$. $SmoothL_1$ can be expressed as  
\begin{equation}
SmoothL_{1} (\Theta) =\frac{1}{N}\sum_{i=1}^M Z^{i}
\label{eq:sid}
\end{equation}
where
 \[
    Z^{i}=
    \left\{
                \begin{array}{ll}
                  0.5\times (DIF)^2   &if |DIF|<1\\
                  |DIF|-0.5           &otherwise,
                \end{array}
           \right.
  \]
and $DIF =HR^{i}_{RGB}-LR^{i}_{MS}$. 

For simplicity, we refer to the $SmoothL_1$ function used here also as $L_1$. Moreover, inspired by Goodfellow's GAN paper \cite{goodfellow2014gan} and the suggested loss function modification to increase derivatives at the beginning of training, we apply a modest modification to improve the $L_{1}$ performance.  More specifically, we choose to minimise the $L1$ cost function in logarithmic scale, that is, we minimise $Log_{10}(L_{1})$. As the cost function approaches zero, following the thought process in \cite{goodfellow2014gan}, using a logarithmic scale loss function could deepen or possibly stretch the cost function surface by changing the minimum possible value from $0$ to $-\infty$. In other words, as opposed to the GAN paper \cite{goodfellow2014gan} where derivatives at the beginning of the training increase, here the derivative around convergence could increase, facilitating further training.

 \subsection{Implementation Details}
 \begin{table*}[t]
\label{table:results}
\centering
 \scalebox{1}{
\begin{tabular}{lc|c|c|c|c|c|c|c|}
\cline{3-7}
                                          &                                & PIRM2018 (SoTA)& \multicolumn{4}{|c|}{Ours} \\ \hline
\multicolumn{1}{|c|}{\bf{Method}}        &\multicolumn{1}{|c|}{\bf{Bicubic}} & RCAN*  & RCAN     & RIRN   & RIRN+Multi-FAN     & RIRN+Multi-FAN$_{log}$  \\
                \hline
\multicolumn{1}{|c|}{\bf{PSNR (dB)}}          &  \multicolumn{1}{|c|}{28.63} & 32.65    & 33.27   &  33.30   & \underline{33.32}    &\bf{33.36}\\ 
\multicolumn{1}{|c|}{} & \multicolumn{1}{|c|}{(3.502)} & (0.01)& (0.01)  & (0.01)   & (0.01)    & (0.01)          \\ \hline
\multicolumn{1}{|c|}{\bf{SSIM}}        & \multicolumn{1}{|c|}{0.4552}    &  0.6367 &  0.6480  & 0.6485    &  \underline{0.6498}    &  \bf{0.6506}   \\ 
\multicolumn{1}{|c|}{} & \multicolumn{1}{|c|}{(0.0691)}&(0.0005)& (0.0005)& (0.0007)& (0.0009)     & (0.0005)      \\ \hline
\end{tabular}}

\caption{PSNR, and SSIM obtained using different models with $g=5$, and $b=3$. Mean and standard deviation (in parenthesis) are calculated between the 11-fold cross-validation experiments.}
\label{table:1}
\end{table*}
Now we specify the implementation details of our proposed Multi-FAN. As mentioned before, the RIRN/RCAN part of our network has $g=5$ RGs. Each RG contains $b=3$ RBs. The kernel size of all our convolutional layers are set to $3\times3$. The convolutional layers in the shallow feature extraction and the body has $C=64$ filters, except at the tail of the RCAN where channels are reduced to $C=1$. The following layers have 64, 32, and 1 channel outputs respectively. Our $L_1$ loss function is applied at the output of RCAN/RIRN block ($SR_{1}$), Multi-FAN ($SR_{2}$) and the final output ($SR_{out}$).

\subsection{Dataset: Generating LR mosaic images}
\label{sec:mosaic_gen}
As explained in the related work, performing SR on demosaiced images gives rise to artifacts related to demosaicing and interpolation \cite{zhou_2018, fu_2018}. For example, SR CNN based methods such as VDSR \cite{kim2016VDSR}, and SRCNN \cite{dong2016SRCNNPAMI} that first interpolate the input LR images up to the scale of the HR images suffer from these artifacts via losing information and decreasing computational efficiency \cite{zhang_2018}. In the more related work of multi-spectral SR \cite{shi_2018,lahoud_2018} (from PIRM2018 spectral SR challenge) SR is carried out on multi-spectral cubes. The images were not interpolated, likely to avoid interpolation artifacts. However, not interpolating the images is likely to have led to the loss of spatial information contained by each wavelength. This is due to the spatial offset of each wavelength in a $4\times 4$ multi-spectral pattern \cite{shoeiby_dataset_2018} (see Figure \ref{fig:mosaic}). In theory, a CNN can learn the geometry of the wavelengths channels using the ground truth images. However, based on our empirical results we show that this is not the case. 

Inspired by the ideas in \cite{zhou_2018, fu_2018}, which all in a way take advantage of the high frequency information of mosaic images, we propose our own procedure which is to generate HR-LR mosaic image pairs from image cubes. As demonstrated in Figure \ref{fig:mosaic} we generate HR and LR mosaic images from HR multi-spectral cubes in the StereoMSI dataset \cite{shoeiby_dataset_2018}, by taking the spatial offset of each wavelength in a multi-spectral pixel into account \cite{shoeiby_dataset_2018}. The HR multi-spectral cubes have a dimension of $14 \times 240 \times 480$, and LR$\times3$ have a dimension of $14 \times 80 \times 160$. According to \cite{shoeiby_dataset_2018}, the multi-spectral pixels have a pattern of $4\times4$, meaning 16 channels. However, two of these channels are redundant, leaving us with 14 channels. We transform this 14 channel cube to its mosaic pattern following the spatial location provided in \cite{shoeiby_dataset_2018}. For the two redundant wavelengths, we assign zero value. In Figure \ref{fig:mosaic}, these two wavelengths shown as black.  The resulting HR and LR mosaic patterns have dimensions $1 \times 960 \times 1920$, and $1 \times 320 \times 640$ respectively.

\section{Experiments}

\subsection{Settings}

\paragraph{Dataset:} To evaluate the effectiveness of our approach, we make use of the PIRM2018 spectral SR challenge dataset~\cite{shoeiby_dataset_2018, shoeiby_pirm2018_method}. The dataset is comprised of 350 multi-spectral pairs of $LR$ and $HR$ images. The training set consists of 300 images plus 30 and 20 images set aside for validation and testing respectively. $HR$ and $LR$ cubes exhibit resolutions of $14\times240\times480$, and $14\times80\times160$ respectively. Following our proposed mosaic generation procedure in Section~\ref{sec:mosaic_gen}, we turn $HR$ cubes to mosaic of size $1\times960\times1920$ and $LR$ cubes to mosaics of size $1\times320\times640$.
\paragraph{Evaluation metrics:}

 The 20 test images were super-resolved to a scale of $\times3$ and evaluated using Pixel Signal to Noise Ratio (PSNR) and Structural Similarity Index (SSIM). For the SSIM metric, a window size of 7 is used, and the metric is calculated for each channel, and then averaged. It is important to note that, while we present SSIM results as it is customary in the SR literature, it is a relatively less descriptive metric in multi-spectral image SR compared to RGB SR. This is due to the fact that the amount of light absorption of a specific material/pixel is mostly related to PSNR rather that a perceptual metric such as SSIM.

\paragraph{11-fold cross-validation:}

To verify the stability of our methods on the PIRM2018 dataset, we carried out an 11-fold cross-validation experiment. The 330 training plus validation images, were randomly divided into 11 folders with one folder iteratively selected as validation and the rest for training. The first fold corresponds to the original dataset division. This cross-validation was only performed for the results presented in Table \ref{table:1}, with model parameters corresponding more closely to the state-of-the-art algorithm from the challenge. Confirming the stability of our methods and obtaining very low standard deviations between the 11 experiments, we continued the rest of our ablation studies with the first fold which corresponds to the dataset in the challenge.

\paragraph{Training settings:}
 During training, we performed data augmentation on our mini-batches of 16 images of our 300 training images, which included random $60\times 60$ cropping of the input image, random rotation by $0^{\circ}$, $90^{\circ}$ ,$180^{\circ}$ , $270^{\circ}$ with $p=0.25$, and random horizontal flip with $p=0.5$. Our model is trained by the ADAM optimizer \cite{adam_2014} with $\beta_1 = 0.9$, $\beta_2=0.999$, and $\epsilon = 10^{-8}$ . The initial learning rate is set to $10^{-4}$ and then halved every 2500 epochs. To implement our models we used PyTorch \cite{pytorch}, and in particular, for our $L_1$ loss function, the $SmoothL_1$ function \cite{smoothl1} was used as the main building block. To test our algorithms, we selected the models with the best performance on the validation dataset, and present the testing results for those models. 
 
\subsection{Effect of using mosaic images}

To evaluate the effectiveness of mosaic images in SR, we compare it with the RCAN trained on mosaic images and original data format as in~\cite{shi_2018}. Note, the data format used in~\cite{shi_2018} does not consider the spatial offset of each multi-spectral pixel. In Table~\ref{table:1}, we provide the results of this comparison as RCAN$^*$ and RCAN. A considerable improvement of $0.62$ in PSNR is achieved only by using our generated mosaic images. We can conclude that, the network is not fully capable of learning the geometrical relationship between the wavelengths (sub-pixels) in a multi-spectral pixel, highlighting the effect of our data representation in improving PSNR and SSIM results.

\subsection{Effect of the CA}
Before we demonstrate the effect of our proposed Multi-FAN module, we first study the effect of CA in the architecture. In our implementation of RCAN, which follows the parameters in \cite{shi_2018} for multi-spectral SR, we have $g=5$ and $b=3$ translating to 15 CA mechanisms. Interestingly, CA seems to deteriorate PSNR performance by $0.03dB$, with results presented in Table \ref{table:1} for RCAN and RIRN (RCAN without CA). We hypothesize that this is due to the fact that our implementation of RCAN is not \textit{very deep}, and CA can help to train very deep networks such as in \cite{zhang_2018}. In fact, according to \cite{zhang_2018}, CA in a very deep RCAN with $g=10$ and $b=20$, 200 CAs can lead to a PSNR improvement of $0.03dB$. Hence, we also train the \textit{very deep} RCAN of \cite{zhang_2018}, with results presented in Table \ref{table:10-20}, showing that CA can improve PSNR by $0.02$ with a \textit{very deep} network. The remainder of the results in Table \ref{table:1}, and \ref{table:10-20} are discussed in the following.

\subsection{Effect of the Multi-FAN module}
\begin{table}
\centering
 \scalebox{1}{
\begin{tabular}{|l|c|c|c|}
\hline
\bf{Metric}       & SR$_1$    & SR$_{2}$          &  SR$_{out}$   \\ \hline
\bf{PSNR (dB)}         &  33.296   & \underline{33.357}& \bf{33.362}   \\
              &  (0.012)  & (0.009)           & (3.698)   \\ \hline 
\bf{SSIM}         &  0.6478   & \underline{0.6502}& \bf{0.6506}   \\    
                  &  (0.0007) & (0.0004)          & (0.0005)\\ \hline 
\end{tabular}}
\caption{Effect of Multi-FAN demonstrated by assessing PSNR and SSIM at three points in our network, output of RIRN (SR$_1$) with $g=5$ and $b=3$, output of Multi-FAN (SR$_2$). These results are averaged over our 11-fold cross validation experiments.}
\label{table:Multi-FAN_study}
\end{table}
\begin{table*}
\centering
 \scalebox{0.9}{
\begin{tabular}{|l|c|c|c|c|c|c|}
\hline
\bf{Method}& RCAN    & RIRN    & RIRN+Multi-FAN & RIRN+Multi-FAN$_{log}$& RCAN+Multi-FAN& RCAN+Multi-FAN$_{log}$  \\  \hline 
\bf{PSNR (dB)}  & 33.39  & 33.37  &  33.39  & \underline{33.42}          & \bf{33.44}  & \bf{33.44} \\ 
           & (3.742) & (3.763) &  (3.791) &(3.790)          &  (3.771)& (3.777)\\ \hline
\bf{SSIM}  & 0.6556  & 0.6535  &  0.6537  & \bf{0.6572}          & 0.6559  & \underline{0.6569} \\ 
           & (0.0614)& (0.0614)&(0.0619)  &(0.06108)        & 0.06147 & (0.0612)\\ \hline 
\end{tabular}
 }
\caption{Mean and standard deviation (in parenthesis) of PSNR, and SSIM obtained using different models with RCAN with $g=10$ and $b=20$ as the baseline.}
\label{table:10-20}
\end{table*}
As explained earlier, the incentive behind our Multi-FAN is to exploit multi-scale semantic features for the task of multi-spectral SR. To quantitatively assess the benefit of the Multi-FAN module, we have trained RIRN+Multi-FAN with results displayed in Table \ref{table:1}. The results are presented in Table \ref{table:1}. An improvement of $0.02dB$ can be achieved by using our proposed Multi-FAN compared to RIRN. The improvement of our RIRN+Multi-FAN compared to RCAN is $0.05dB$. To study the effect of Multi-FAN in more detail,  in Table \ref{table:Multi-FAN_study}, we have provided PSNR and SSIM results for the image at the output of RIRN (SR$_1$), at the output of Multi-FAN module (SR$_2$), and the output of RIRN+Multi-FAN (SR$_{out}$). Note that the results obtained via the Multi-FAN module are superior to the results at the output of RIRN. Besides, the final results at SR$_{out}$ is superior to both results at SR$_1$, and SR$_2$, showing that the two networks (Multi-FAN, and RIRN) are learning complementary information. 

When we changed our $L_1$ loss to $Log_{10}(L_1)$, the results improved by another $0.04dB$. In comparison to the RCAN baseline, our Multi-FAN module and logarithmic loss improve the PSNR results by $0.09dB$. Note that due to recent progress in CNN based SR, in particular, the contribution of the RCAN paper \cite{zhang_2018}, further improvement in PSNR is a substantially challenging task. Nevertheless, our $0.09$ improvement in PSNR in dB, which is purely due to the architectural contribution and logarithmic loss,  amounts to a $\%2.09$ improvement in ratio. Overall, with our mosaic data representation, Multi-FAN module, and logarithmic loss, a generous PSNR improvement of $0.71dB$ is achieved. Also, note that the very small PSNR and SSIM standard deviation of our 11-fold cross-validation experiment in Table \ref{table:1} is evident of reliability of our results and stability of our method. 

In comparison to the top PIRM2018 algorithm, our algorithm clearly outperforms it. No self-ensemble algorithm was used in the post-processing stage to achieve further improvements. These results purely demonstrate the effect of our data-representation, our network architecture, as well as our logarithmic loss function contributions.

For the sake of completeness, we also further investigated the effect of Multi-FAN and logarithm loss on the \textit{very deep} RCAN in \cite{zhang_2018} with $g=10$, and $b=20$ with results presented in Table \ref{table:10-20}. Since the results in Table \ref{table:1} exhibits very low standard deviations between the 11-fold cross-validation experiments, here, and for the rest of the paper, the experiments are carried out only on the first fold. Looking at the results in Table \ref{table:10-20}, the Multi-FAN module has resulted in a $0.02 dB$ improvement in PSNR when used with RIRN, and an improvement of $0.05dB$ in PSNR when used with RCAN. Logarithmic loss has again shown to be effective in improving the results.


\subsection{Effect of different loss combinations and their logarithmic versions}
\begin{table}[h]
\centering
 \scalebox{0.95}{
\label{table:time}
\begin{tabular}{|l|c|c|c|c|}
\hline
\bf{Loss}       & SSIM      & SSIM$_{log}$  & L1 +      &  L1$_{log}$+ \\ 
\bf{function}   &           &               & SSIM       & SSIM$_{log}$\\ \hline
\bf{PSNR (dB)}       &  33.256   &\underline{33.27}&   33.26  & \bf{33.28}       \\
            &  (3.69)  & (3.68)      &   (3.69) & (3.69)\\ \hline 
\bf{SSIM}       &  0.6567   &  \bf{0.6575}  &   \underline{0.6570}  & 0.6554   \\    
                &  (0.0606) & (0.0609)      &  (0.0610) & (0.0607)\\ \hline 
\end{tabular}}
\caption{Effect of logarithmic loss for different loss functions for RCAN with $g=5$ and $b=3$.}
\label{table:log_study}
\end{table}
To further validate the effectiveness of the logarithmic version of a loss function, and also to investigate the effect of different loss functions on training, we carried out the following experiments. We train our RIRN+Multi-FAN with 1. SSIM loss function and its logarithmic version, 2. the summation of SmoothL1 and SSIM loss functions, and 3. the summation of the logarithmic versions of both loss functions. The results are presented in Table \ref{table:log_study}. From these results, it is apparent that a logarithmic loss improves performance both on PSNR and SSIM. Our best PSNR is achieved when using the logarithmic version of both L1 and SSIM loss functions. Our best SSIM result is obtained when using the logarithmic version of the SSIM loss function on its own.

\subsection{Qualitative results}
\begin{figure*}
\centering
  \includegraphics[trim={2.5cm 11.5cm  1.5cm  11.5cm },clip,width=\linewidth]{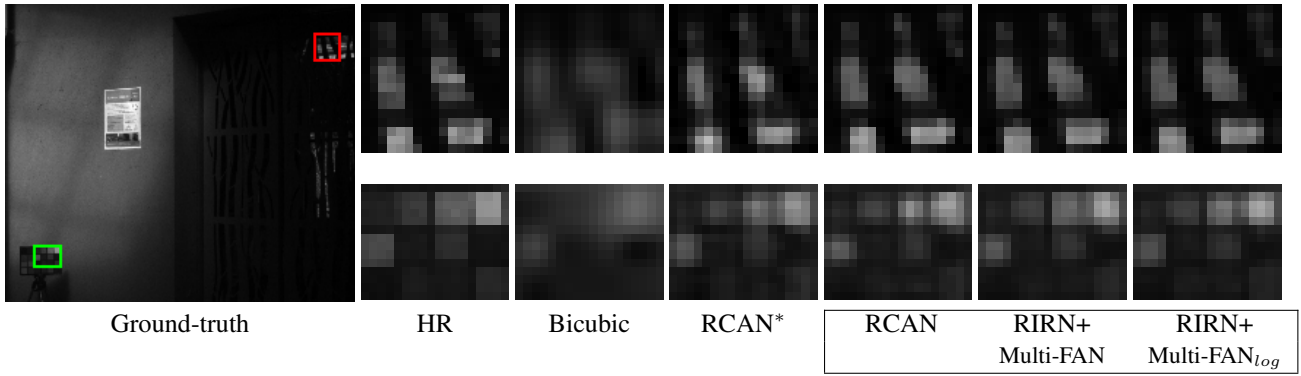}
  \caption{Qualitative results at $\lambda=612.9nm$.   RCAN* indicates the approach from PIRM2018 challenge. The bounding box contains, from left to right, the results demonstrating the effect of our data representation, our Multi-FAN network and logarithmic scale loss function respectively.}
  \label{fig:im1}
\end{figure*}
\begin{figure*}
\centering
  \includegraphics[trim={2.5cm 11.9cm  1.8cm  12cm },clip,width=\linewidth]{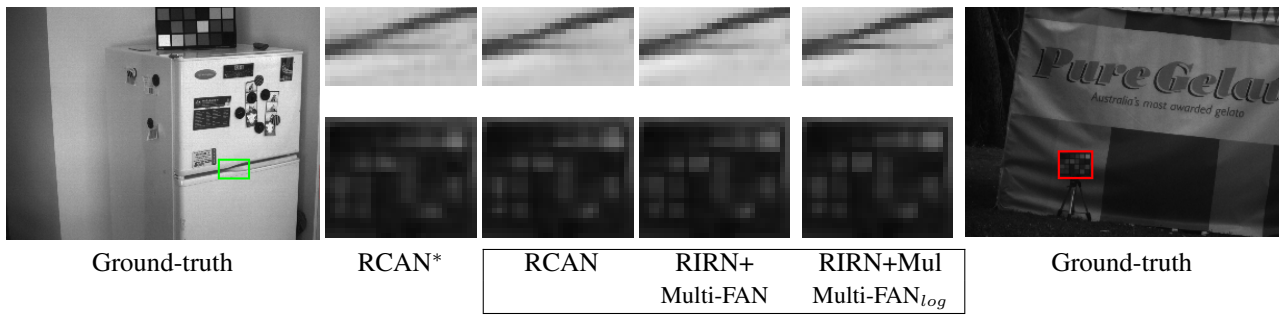}
  \caption{Qualitative results at $\lambda=562.5nm$ (left), and $\lambda=577.3nm$ (right).   RCAN* indicates the approach from the PIRM2018 challenge. The bounding box contains, from left to right, the results demonstrating the effect of our data representation, our Multi-FAN network and logarithmic scale loss function respectively.}
  \label{fig:im2}
\end{figure*}

The qualitative results are presented in Figure \ref{fig:im1} and \ref{fig:im2} displaying different wavelength channels. The images are from the test dataset and belong to wavelength channel $612.9nm$ in Figure \ref{fig:im1}, and wavelength channels $577.3$ and $562.5$ in Figure \ref{fig:im2} respectively. The improvements introduced via mosaic representation can be seen in both Figures by comparing RCAN* results to RCAN. In all images, the edges are better defined. This further improves, yet more subtly, when we introduce our MFA-FAN and logarithmic version of the loss function. This effect can be better qualified by assessing the edges in the colour checker. Note, our ground truth images, due to the low spatial resolution of multi-spectral cameras, do not enjoy the high resolution of ground truth images usually used in the RGB SR literature \cite{huang2015single,fujimoto2016manga109,bevilacqua2012low,martin2001database}. Besides, due to the grayscale nature of the images, visual improvements regarding the colours (intensity of the images) are more difficult to qualify.


\begin{table}
\centering
 \scalebox{0.7}{
\label{table:time}
\begin{tabular}{|l|c|c|c|c|c|c|}
\hline
                 & \multicolumn{3}{|c|}{$g=5$, and $b=3$}& \multicolumn{3}{|c|}{$g=10$, and $b=20$} \\
                 \cline{2-7}
\bf{Method}      & RCAN    & RIRN   & RIRN +& RCAN &RIRN & RIRN + \\     
                 &         &        &  Multi-FAN  &     &       & Multi-FAN  \\ \hline     
\bf{Params}  & 1.37M   & 1.36M  & 1.54M   & 15.33M   & 15.21M   & 15.57M\\ \hline 
\bf{Time (sec)}  & 7$e^{-3}$ & 3$e^{-3}$& 4$e^{-3}$ & 2.5  & 0.35 & 0.46\\ \hline 
\bf{PSNR(dB)}  & 33.27 & 33.30 & 33.32 & 33.39  & 33.37  & 33.39\\ \hline %
\end{tabular}}
\caption{Effect of CA and Multi-FAN on inference time (seconds), number of parameters, and PSNR (dB).}
\label{table:time}
\end{table}
\subsection{Inference time:}

Since CA involves multiplication of a calculated weight vector by the input feature map, it can be computationally relatively expensive if used in abundance. In situations where one is forced to compromise between speed and accuracy, our Multi-FAN module can be exploited to improve performance while still maintaining a short inference time. To demonstrate this, Table \ref{table:time} presents number of parameters, inference time, and PSNR results for the same loss function (SmoothL1) for RCAN, RIRN and RIRN+Multi-FAN for the two different network sizes discussed in this paper, that is, one with $g=5$ and $b=3$ \cite{shi_2018}, and the other with $g=10$ and $b=20$ \cite{zhang_2018}. 

In the case where $g=5$ and $b=3$, we have 15 CA mechanisms. Removing these 15 CAs and instead employing a Multi-FAN module not only improves the performance in terms of the fidelity of images but also reduces inference time as well. Due to the small values of inference time for this scenario, we averaged the inference times between our 11-fold cross-validation experiments. In the case of a \textit{very deep} residual network with $g=10$ and $b=20$, we have 200 CA mechanisms. Removing these causes a $0.02dB$ reduction in PSNR. However, when using a Multi-FAN module a significant $\times 5.5$ reduction in inference time is achieved while still maintaining its PSNR performance.
\section{Conclusions}
In this paper, we proposed a new architecture for the task of super-resolution of multi-spectral images. Experimental results show the effectiveness of our approach as well as the data format we generate. In particular, we show that 1) incorporating features at multiple semantic levels (e.g., from different RGs) is beneficial and 2) our generated mosaic images are more effective than the original raw multi-spectral cubes. Quantitatively, this work improves on  \cite{shoeiby_dataset_2018} by a total of $0.71dB$ in PSNR with $0.62dB$ improvement being due to mosaic pattern generation, and $0.09dB$ was due to our proposed Multi-FAN network and our choice of logarithmic scale loss function.

Regarding inference time, we showed that removing CA mechanisms and using the Multi-FAN network instead can result in significant improvement of inference time, \eg a $\times 5.5$ decrease for the \textit{very deep} RCAN \cite{zhang_2018} with no compromise in PSNR results. 

{\small
\bibliographystyle{ieee}
\bibliography{references}
}

\end{document}